# Dynamical issues in interactive representation of physical objects

Jean-Loup Florens*
ACROE, Grenoble

Alina Voda†
Laboratoire d'Automatique
de Grenoble

Daniela Urma‡
ICA, Grenoble

**ABSTRACT**

The quality of a simulator equipped with a haptic interface is given by the dynamical properties of its components: haptic interface, simulator and control system. Some application areas of such kind of simulator like musical synthesis, animation or more general, instrumental art have specific requirements as for the "haptic rendering" of small movements that go beyond the usual haptic interfaces allow. Object properties variability and different situations of object combination represent important aspects of such type of application which makes that the user can be interested as much in the restitution of certain global properties of an entire object domain as in the restitution of properties that are specific to an isolate object.

In the traditional approaches, the usual criteria are founded on the paradigm of transparency and are related to the impedance error introduced by the technical aspects of the system.

As a general aim, rather than to minimize these effects, we look to characterize them by physical metaphors conferring to haptic medium the role of a tool. This positioning leads to firstly analyze the natural human object interaction as a simplified evolutive system and then considers its synthesis in the case of the interactive physical simulation. By means of a frequential method, this approach is presented for some elementary configurations of the simulator

## 1 INTRODUCTION

As underlined in several publications, the haptic device domains have quickly grown during the two past decades [1], [2], [3]. Most known applications belong to the Virtual Reality (VR), domain that reached an important place, extending classical industrial manipulators and in which the haptic display appears as complementary of the visual one. This development is widely stimulated by the increase of computer power and the ability to compute in real time more and more realistic and complex environments. Several investigated questions concerning dynamic properties in haptic applications refer mainly to the stability and the quality of the haptic rendering. Early works [4], [5], [6] on these questions have been carried on in the context of industrial or scientific applications of tele-operation and master slave systems.

Later on, in the new VR domain, haptic devices has been integrated to work in conjunction with a simulation process. In many cases this later domain is issued from the earlier, the virtual environement being considered as equivalent of the remote tele-operated one.

However, haptic is involved in a less-known domain, the multisensory simulation for instrumental arts, especially music and moving image synthesis Historically, this specific artistic field comes from sound synthesis discipline and acoustics [7], [8], [9], thus has not completely appeared in the context of the most known Virtual Reality or teleoperation in robotic applications.

Aside the previous domains, the multisensory application domain has primarily focused its studies and technical efforts on the dynamic quality [10], [11], [12], [13]. instead of shape and space rendering.

Although many technical solutions concerning haptic interfacing could be shared between these different domains, the needs are not completely the same and some specific methodologies have appeared in the domain of multisensory simulation for arts, that point different approaches concerning the dynamic issues.

The topic developed here relates to one of these issues, and involves the lowest level of the human-object interaction. The methodology is based on an elementary frequential approach in which we try to consider the variations induced by the technical device as part of the modeled universe. This provides simple means for a global characterization of the technical effects as a deformation of the parameter space.

## 2 SUMMARIZED STATE OF THE ART ON DYNAMIC SYSTEMS APPROACHED

Haptic control design appears as a particular complex domain of the dynamic systems control. It presents similar difficulties with robotics, from combination of complex motion with control issues, to load variability and uncertainty [14]. Moreover, in the haptic domain the human presence in the loop leads to unusual operating criteria due to his sensitivity to some technical imperfections such as backlash in the transmission links or the actuator noise.

### 2.1 Dynamic issues in tele-operation systems.

Studies on haptic interface control have started focusing on problems related to the stability of tele-operation chains [4], [5], [6]. The sources of such system instabilities have been investigated through the variation of mechanical impedances (contacts with stiff objects) and in the context

This work has been supported by the French Ministry of Culture and by the FP6 Network of Excellence IST-2002-002114 - Enactive Interfaces.
*ACROE, INPG, 46 Av. Felix Viallet 38000 Grenoble, France, florens@imag.fr
†*Laboratoire d'Automatique de Grenoble.* Alina.Voda@inpg.fr
‡*ICA*, INPG, 46 Av. Felix Viallet, 38000 Grenoble, France,

of the variability of the interaction between human operator and a stable environment (passive or not).

These studies have been highly intensified during the 90's [15], [16], [17].for which the passivity method has been mostly used to analyze the stability of teleoperation systems and to design adequate haptic controllers. Thus, the stress has been put on checking the passivity of the components in interaction: manipulated object, human operator, and the link between these two elements.

Such works mainly refer to the teleoperation scheme as a two port coupling network [17], [18], [19]. This scheme allows a clear definition of performance measures, such as *transparency*.

Transparency requires that the impedance transmitted to the operator is equal to the remote object impedance transmitted to the operator through the haptic system [20].

However, the degree of stability and the degree of transparency act as two antagonist properties. Reaching a high transparency implies to increase the gains and hence to reduce the stability margin [20].

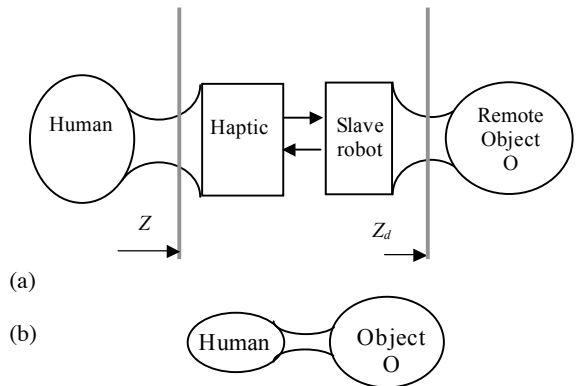

Figure 1 The transparency of the tele-operation chain (a) is defined as the impedances equality :
This condition refers to the direct natural interaction with a same object (b)

The concept of passivity has been applied to this two-port model of teleoperator to insure stability conditions whatever the load impedances at the two ends. Many of these studies were focused on the delay, since it is a problem frequently experienced in long distance or through complex network teleoperation links.

### 2.2 Dynamic issues extended to simulators

The literature treating the stability problems in VR systems has been developed lately [21], [22], [23], [24], [25]. In the simulator case, even if the methods to analyze stability are similar to those used in tele-operation, specific features have to be considered. Since the object is simulated, its dynamic representation as difference equations may introduce approximation errors. Consequently, the number of possible sources of instabilities is extended.

Similarly, the transparency definition for simulator systems relies on impedance equality: the virtual environement impedance one side and the impedance brought out to the user by the haptic interface. This definition leads to design controllers for minimizing the corresponding impedance error as described in [23], [24]

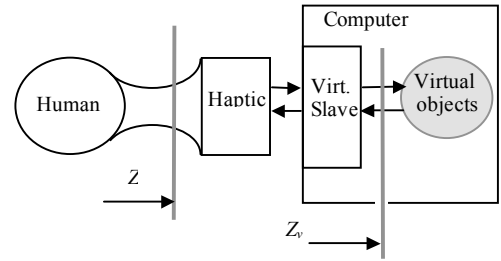

Figure 2 Transparency in the case of simulation.

. Compared to the tele-operation case the simulator transparency paradigm presents an additional difficulty since the relationships linking the real and the virtual physical variables are, for a part, arbitrary or depend on a reference interface. This point concerns in particular the D/A & A/D conversion operators.

As a general rule transparency is reachable only in a limited frequency bandwidth due to unavoidable effects coming from the simulation support and from the haptic device.

Our aim is to extend the transparency criterium with a frequency calibration keeping into account the previous unavoidable effects. To achieve this, these effects should be modeled and represented as parts of the virtual environment. Such an extended transparency criterium would better manage the performance/stability compromise.

The present study is a first step towards the definition and validation of such an extended transparency criteria.

### 3 PARAMETER SPACE CONSISTENCY CONSTITUTING THE PARAMETER SCALES

An important concern in the described application is not only the quality of rendering for some given singular objects, but it has to satisfy some realistic criteria concerning the parameter scales, thus providing consistency on the whole field of possible designable objects. These rules are related to the physical parameter scales and the different ways in which the user can apprehend these parameters, particularly:

1- The quantitative knowledge on the model, for instance how stiff is a given element or part of the object

2- The qualitative knowledge, for instance how is made this object or which combination of components constitutes that given stiff part

3- The multisensory interaction with the object through gesture, sound and vision. In this situation, the physical features (i.e. the stiffness) indicate different types of phenomena that address not only static effects (related to the determination of an equilibrium), but also mainly dynamical effects related to temporal properties of motions induced by the interaction.

Considered only from the virtual universe side, the parameter values possess their own consistency in relation with the assembling operation. For example, an assembly of two springs of stiffness K1 and K2 provides the same behavior than a single (K1+K2) spring, whatever these stiffness are used for in simulation. In other words the qualitative definition of the virtual object and the quantitative one are consistent. Consequently we can assume that the virtual universe as well as the real one presents separately their own consistency.

However the consistency problem appears when we consider the third category of objects (the 1rst and 2nd being respectively the real objects and the virtual objects) that are issued of the hybrid composition of virtual and real physical systems across the haptic interface.

To clarify this point we can consider the experiment described in the following (Table1). In this proposed example the simulation system is supposed to be very simple. In particular the computation delays are supposed to be null and no holder are employed. The analysis is then very easy and shows more clearly the concerned effect.

A same real mass m is coupled once with a real spring k and once with a virtual spring K (Table 1, I and II). The K parameter is then tuned to provide an equivalent behavior of both mass-spring systems i.e. a same eigen-frequency. The K virtual spring is then supposed to represent the k real spring relatively to the described situation.

| (I) Real mass coupled with real spring k. 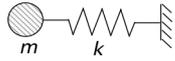 | (II) Real mass coupled with virtual spring K 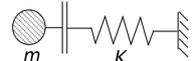 |
|---|---|
| (III) Real mass coupled with real springs, 2k 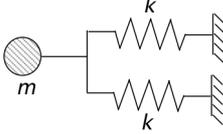 | (IV) Real mass coupled with virtual springs, 2K 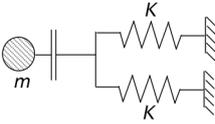 |
| | (V) Real mass coupled with virtual spring K and a real spring k 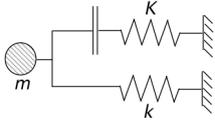 |

Table 1. Different coupling over the haptic interface induce some parameter inconsistencies. The double vertical line represents the borderline between real and virtual worlds.

The systems I and II are given equivalent. The systems III, IV, and V are not equivalent.

If we consider now the assembly of several k springs instead of one (Table 1, III) and the representative virtual assembly of several K virtual springs (Table 1, IV), we will observe a difference between the frequential behavior of the two systems. By bunching the same real mass m with the hybrid assembly of the real and virtual springs (Table 1, V), we exhibit a new distinct behavior.

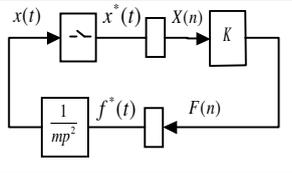

$$\frac{m}{T^2}(1-z)^2 + Kz = 0 \quad (1)$$

$$K = \frac{m}{T^2}(2 - 2\cos T\sqrt{\frac{k}{m}}) \quad (2)$$

Table 2. Minimal simulation model realizing the real mass /virtual spring assembly. The simpler case corresponds to a no-delay ideal simulation that is described by the z characteristic 2nd order equation (1). In this case the tuning relation that provides K from k is given by (2).

These behavioral differences are due to the combined effects of haptic interface and signal treatment support used for simulation (limited bandwidth and discrete time representation).

## 4  THE LOW LEVEL OF HUMAN OBJECT INTERACTION. "T.H.S." HYPOTHESIS.

The field of interest of instrumental interaction contains many situations in which the man-object coupling is the seat of phenomena to broad spectrum [33], [34], [35]. It actually appears in most current human-object interactions like grasping, colliding, exerting a friction on a rough surface, as soon as the object is "shaped" and presents a sufficient stiffness. In the particular case of some sound-producing objects like elementary musical instruments, the sounding effect are completely tighten to the high frequencies phenomena that appear in the temporary system. that results from the hand-object interaction.

In many similar conditions of interaction, we can consider the quasi-autonomous physical system constituted by a limited part of the human mechanics and a part belonging to the object, both situated near the human-object borderline. By its eigen-motion, this system is able to produce, during short amounts of time, the observed high frequency phenomena. We call it THS, temporary hybrid system (Figure 3), and we reserve the term of "object" to what is completely external to the human.

The THS is an evolutive system whose physical parameter are the consequence of various combinations of the human and objects dynamics. The primary type of THS is the elementary oscillator in which we suppose that the visco-elastic part is inside the object and that the human-object borderline is inside the THS mass. This type of THS is not supposed to be the best hypothesis for all the situations. It is just the simpler to study with respect to the structure of the simulator interface (impedance mode). Other significant case studies can be outlined following this principle.

To clarify the simulation situation, we will assume that the THS is distributed as follows: (1) the mass element is composed of the human part and of the material part of the haptic interface, (2) the viscous element is composed of the estimated real linear friction of the HI and of a virtual viscosity, the elastic part being only a virtual spring.

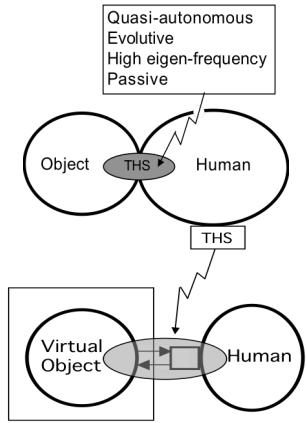

Figure 3. Temporary hybrid system:

(a) in the natural situation;
(b) supported by the interactive simulator

The intended method consists in comparing two closed THS systems, in the real and in the simulation situations. This could be an equivalent of the criterion applied to the open system (the 2 port network).

Finally, according to these assumptions, the main difference from the classical approaches is that the analyzed system is not the manipulated object in itself but the chosen THS. In addition, the criteria of analysis concern not only some particular sets of parameters of the THS, but the whole family of possible parameter sets that the THS may get along its variations.

Following this hypothesis, the primary role of an interactive simulator and its haptic interface is to carry on similar THS. Moreover, if a computed object Oc is the virtual representation of a given object O, the set of temporary systems generated by the interaction of the human with Oc should be the closest to the set generated by human interaction with O, for all the configurations of THS that the interaction may generate. The concepts underlined in the section 4 directly deal with this criterion with respect to the THS. In the following section, we present firstly the related analysis method and then we will discuss on some primary simulator configurations.

## 5 ANALYZING THE PARAMETER SPACE DEFORMATION

Since the involved elementary system consists in a simple linear oscillator (Mass, Spring, Damper), the analysis could be related to the transformation of the characteristic constants, i.e. its eigen-frequency and damping constant. However, instead of directly displaying these deformations in the z or p plane, we will track the impact of the simulator configuration on the structural properties of the object. Indeed, if we call K the stiffness and B the damping parameters of the hybrid oscillator (for example the THS in the situation of interactive simulation), we can generally define the k stiffness and b damping parameters of a real oscillator that presents the same eigen-frequency/damping. Thus, for a given simulator, we can define the deformation of its physical parameter field as the corresponding transformation (K,B) -> (k,b).

A graphical representation of this transformation consists in drawing the iso k and iso b curves in the K,B plan. This curves network can be used as new coordinate system that provides the k,b values that correspond to a given K,B point. The algorithm to obtain these curves simply consists in sweeping the k,b domain and solving the KB parameter equation defined from the characteristic equation at each value of z deduced from the eigen-frequency of the reference k,b system. The diagram of this algorithm is shown Figure 4.

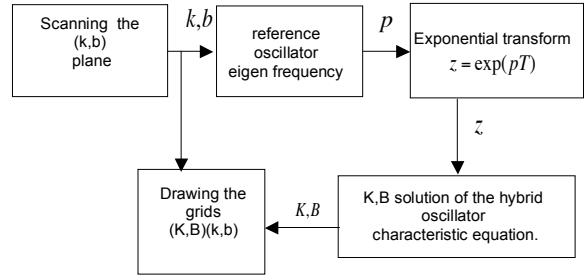

Figure 4. Algorithm scheme for drawing the k,b curves in the K,B plane.

The obtained curvature of the grid represents the degree of deformation introduced on the parameter scales. The orthogonal shapes represent the identical transformation (ideal case). Two types of typical local deformations are possible: (1) The homothetic deformation in which the curves families remain orthogonal but the step sizes are changed. In this case the two scales are independently transformed. (2) The loose of orthogonality between the two families that indicate a coupling between the two scales K and B (i.e. modifying only the K parameter varies also the damping).

## 6 ANALYSING DIFFERENT SIMULATOR IMPLEMENTATIONS.

The retained cases in study refers to different implementation of an haptic simulator working in the open loop impedance mode configuration. [26],[28],[29].

The two basic components of this simulator are shown in Figure 4.

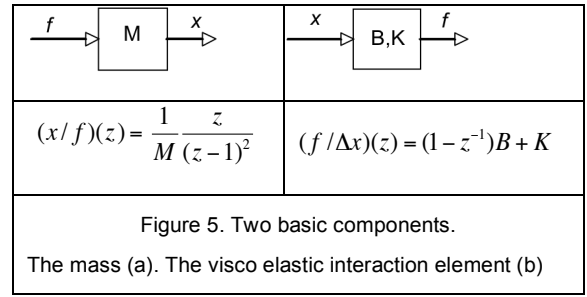

Figure 5. Two basic components.
The mass (a). The visco elastic interaction element (b)

The mass element is based on the usual 2nd order centered derivation scheme and the vico-elastic element is based on the first order Euler derivation scheme. In the studied following example only the visco-elastic element is used.

The first case corresponds to the un-realistic null delay situation.. The second example is a more realistic situation in which the computation delay is taken in account. In the third case, the effect obtained by introducing a real damping which is a known method to improve the stability [15], [36] is analysed.

The general diagram corresponding to the three studied cases is shown in Figure 6.

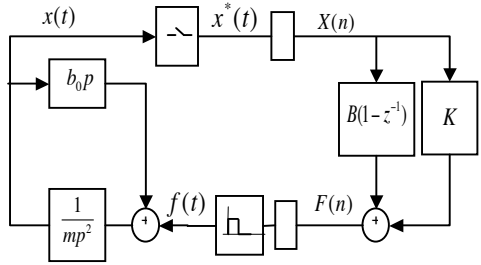

Figure 6 Haptic simulator model.

In these different cases the characteristic equation of the system is assumed to be of the form:

$$P(z)K + Q(z)B + R(z) = 0 \qquad (1)$$

where P,Q,R are polynomial fractions or polynoms characterizing each system.

### 6.1 No delay simulator: Analyzing the numerical scheme effects.

This case corresponds to an idealized (and unrealistic) haptic interfacing in which the sampling presents no integration at the input and no zero order holder at output and the computation is supposed to introduce no delay.

The system is then equivalent to a digital oscillator made of the two elements represented Figure 5. Indeed it is easy to show that in this case, the sampled image of the real mass is identical to a virtual mass element as defined in Figure 5a

The corresponding P,Q,R are then :

$$P(z) = z \quad Q(z) = z-1 \quad R(z) = (z-1)^2 \qquad (2)$$

The analysis in this case exhibits the effects of the numerical schemes and in particular the effect of the 1rst order Euler scheme that is used in the virtual damping element.

The results presented in Figure 7 is a dispersion of the K scale while B scale is not influenced by the variation of K. This first situation is interesting because it shows that although the haptic interface is idealized important parameter distortions appear. These distorsions are the consequence of the discrete time representation.

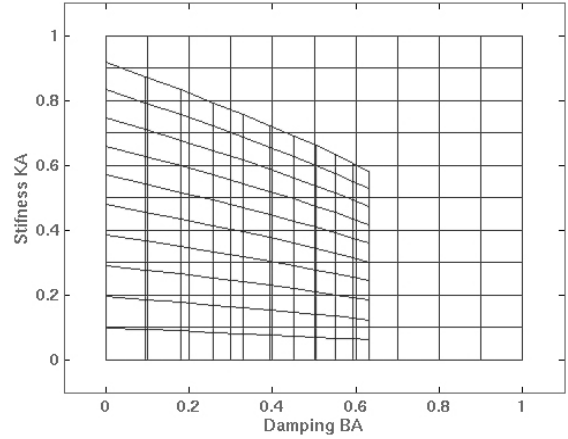

Figure 7. The effects of discrete time representation with an ideal (no delay) haptic interface

### 6.2 Introducing the computation delays

In this case the considered system is defined by:

$$P = z \qquad Q = z-1 \qquad R = z(z-1)^2 \qquad (3)$$

This is the same system as previous with 1 sample delay that corresponds to the minimal computation time. Compared to the previous case the transformation creates important parameter distortions (Figure 8). The stability limit in the K,B domain corresponds to the b=0 curve that presents the known shape and limits the domain of usable K,B parameters. Inside this domain the K,B parameters represent no longer independently the stiffness and damping properties. One more important drawback that can be noticed from Figure 8 is that the asymptotic behavior of the transformation for the low values of K,B is not the identity, contrary to the previous case : the grid is not tangent to the orthogonal reference grid, that shows also that increasing the sampling and computation frequency cannot improve the quality of the system

We can remark that we have not taken in account the supplementary delay normally introduced by the zero order holder and the 1 sample integrator at input. Even in the simpler case of ideal sampling the effects introduced by one sample delay are predominant on the previous type of distorsions.

The delays are generally considered as a source of instability and corrected according to this only criterion. The proposed representation shows that the stability issue is a particular case of high distortion of the parameter scales. Various known ways to (partially) overcome the drawbacks due to delays should be analysed in order to evaluate their influence on the parameter scales transformation. We limit the presentation to one of the most known that is related to the real damping of the haptic device.

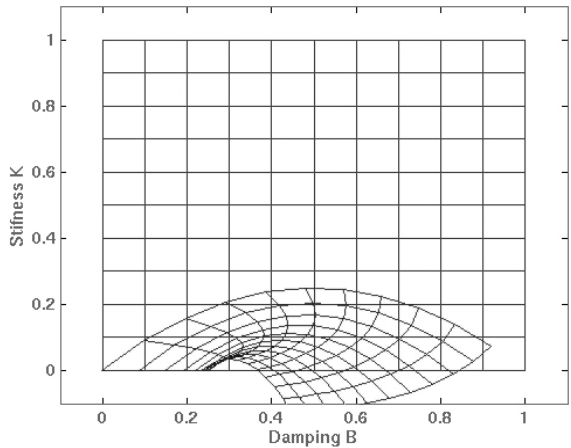

Figure 8. The combined effects of discrete time representation and computation delays

### 6.3 Introducing a real damping.

This useful effect has been shown by Colgate [15] as a mean to improve the stability and K limit in the impedance mode. The introduction of real damping $b_0$ leads to the following formulas :

$$P = z \qquad Q = z-1 \qquad R = z(z-1)(z - e^{-b_0})\frac{b_0}{1 - e^{-b_0}} \qquad (4)$$

The Figure 9 represents the (k,b) grid in this case with a real damping value b0=0,5 (normalized units) in which the $b_0$ bias is taken in account in the definition of the represented B scale.

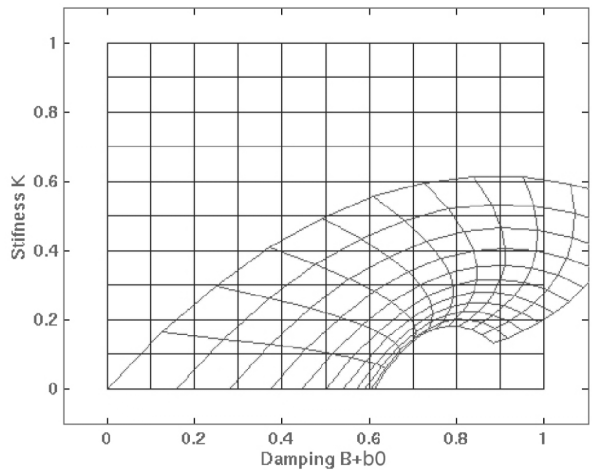

Figure 9. The combination of previous effects and real damping (b0) on parameter scales. The considered virtual B scale takes the b0 bias in account

This configuration presents significant improvements concerning the parameter transformation: the stable area in K,B plan, the compression along the b scale. However the asymptotic behavior at low values of K,B is no longer the identical transformation and high distortions on the parameter scales remain. This existing parameter distortion illustrates that the influence of the delay still exist in a corrected system.

Many other simple correction means could be evaluated from this frequential approach. In conjunction with the real damping the use of additional speed and acceleration sensor may be used to compensate the effect of delay.

## 7  CONCLUSION

We have considered the low level of the haptic interaction and defined some criteria relative to the correctness at this level of an interactive simulator.

This criterion is based on temporary hybrid system modeling of the haptic interaction and we have situated the main differences with the more usual criteria that are based on the two port model. We have equally presented the field of interest of such a criterion.

We have presented an elementary frequential analysis method and a corresponding graphical representation.

This tool corresponds to a need in the domain of multisensory simulation for instrumental art in which the quality relies more on the global rendering of the variety of object than of the objects taken separately. It situates the question of stability and the question of performance at the same level.

This tool focuses the analysis on the very primary level of the interaction and allows the definition of a method to evaluate technical dispositions and corrections for improving the behavior of the simulation at this level.

Further work concern first experimental evaluations of the THS hypothesis and the effects of k,b distortion corrections on simple impedance mode controlled systems. Then the method will be extended to other types of haptic control modes.